\documentclass[sigconf]{acmart} 
\renewcommand\footnotetextcopyrightpermission[1]{} 
\usepackage{kotex}
\usepackage{algorithm}
\usepackage[noend]{algpseudocode}

\AtBeginDocument{%
  }

\begin{document}

\title{Node Embedding for Homophilous Graphs with ARGEW: Augmentation of Random walks by Graph Edge Weights}


\author{Jun Hee Kim}
\affiliation{%
  \institution{NCSOFT}  
  \country{Republic of Korea}
}
\email{junheekim@ncsoft.com}

\author{Jaeman Son}
\affiliation{%
  \institution{NCSOFT}  
  \country{Republic of Korea}
}
\email{jaemanson@ncsoft.com}

\author{Hyunsoo Kim}
\affiliation{%
  \institution{NCSOFT}  
  \country{Republic of Korea}
}
\email{aitch25@ncsoft.com}

\author{Eunjo Lee}
\affiliation{%
  \institution{NCSOFT}  
  \country{Republic of Korea}
}
\email{gimmesilver@ncsoft.com}


\begin{abstract}
Representing nodes in a network as dense vectors i.e. \textit{node embeddings} is important for understanding a given network and solving many downstream tasks. 
In particular, for weighted homophilous graphs where similar nodes are connected with larger edge weights, we desire node embeddings where node pairs with strong weights have closer embeddings. Although random walk based node embedding methods like node2vec and node2vec+ do work for weighted networks via including edge weights in the walk transition probabilities, our experiments show that the embedding result does not adequately reflect edge weights. In this paper, we propose ARGEW (Augmentation of Random walks by Graph Edge Weights), a novel augmentation method for random walks that expands the corpus in such a way that nodes with larger edge weights end up with closer embeddings. ARGEW can work with any random walk based node embedding method, because it is independent of the random sampling strategy itself and works on top of the already-performed walks. With several real-world networks, we demonstrate that with ARGEW, compared to not using it, the desired pattern that node pairs with larger edge weights have closer embeddings is much clearer. We also examine ARGEW's performance in node classification: node2vec with ARGEW outperforms pure node2vec and is not sensitive to hyperparameters (i.e. consistently good). In fact, it achieves similarly good results as supervised GCN, even without any node feature or label information during training. Finally, we explain why ARGEW works consistently well by exploring the coappearance distributions using a synthetic graph with clear structural roles.
\end{abstract}





\maketitle
\pagestyle{plain}

\section{Introduction}

Networks, which are defined by entities and their connections, contain important information, such as relationships and roles of entities, that tabular or sequential types of data do not. In order to extract such significant information from networks, each node needs to be properly represented. Hence one of the tasks that has been extensively researched in relation to network data is embedding nodes. 
For each node in a given network, we want to obtain a real-valued vector that reflects information while taking into account the network structures. 

The notion of ``node similarity'' depends on the context and can be expressed in several forms such as having similar node attributes (e.g. paper's bag-of-words vector in a citation network), or the same node label (e.g. paper's topic). Naturally, one would desire node embeddings where similar nodes have close embedding vectors, while dissimilar nodes locate far away in the embedding space.

One of the common properties of real-world networks is \textit{homophily}, where similar nodes more tend to have edges compared to dissimilar nodes \citep{platonov2022characterizing}. For instance, in a friendship network, people with similar hobbies might easily become friends. Further, for a homophilous graph with edge weights, two nodes that are \textit{more} similar will naturally have \textit{larger} edge weights. Hence, for weighted homophilous networks, the desired properties of node embeddings can be illustrated in two aspects: 1) node pairs with an edge have closer embeddings compared to those without an edge, and 2) node pairs with large-weight edges tend to have closer embeddings compared to those with small-weight edges. Note that if we view a node pair with no edge as a zero-weight edge, then those two properties can just be stated as the second one alone. Random walk based node embedding methods, such as node2vec~\citep{grover2016node2vec} and node2vec+~\citep{liu2023accurately}, do work for weighted networks via including edge weights in the walk's transition probabilities. Our experiments on weighted homophilous networks, however, show that the node embedding results do not satisfy those desired properties enough. 

In this paper, we propose a data augmentation technique for random walk based node embeddings, which aims to generate embeddings that satisfy the aforementioned two desired properties for weighted homophilous networks. Our method can be used with any random walk based node embedding method, because it is independent of the walk sampling strategy itself and works on top of the already-performed walks. To the best of our knowledge, there has not been any node embedding approach that augments the random walks themselves to reflect edge weights to the embeddings. We demonstrate how our method affects the embeddings by comparing the results of the same walk strategy with and without our method. Furthermore, in order to demonstrate the practical usefulness of our method, we use a network constructed using log data of a massively multiplayer online role-playing game (MMORPG) called \textit{Lineage W}\footnote{https://lineagew.plaync.com/} in addition to benchmark network datasets.

The remaining paper is organized as follows: The next section explains some of the related works about node embedding, especially random walk based techniques. Section 3 introduces our proposed method, and Section 4 demonstrates our experimental results. Lastly in Section 5, we summarize our work and discuss some limitations and potential future work regarding our paper.
\section{Related Work}

Given network data, one would naturally seek node-level features that describe the characteristics of each node. These can further be used for node-level tasks such as node classification directly or even edge-level and graph-level tasks by being combined somehow to the desired level. Several types of methods for constructing node features have been proposed. 

As explained by \citet{grover2016node2vec}, methods that utilize network properties exist \citep{gallagher2008leveraging,henderson2011s} but require hand-engineering efforts, and methods based on dimensionality reduction of network-related matrices \citep{belkin2001laplacian,roweis2000nonlinear,tenenbaum2000global,yan2006graph} struggle from computation cost and performance limitations. Another type of methods creates node embeddings by performing random walks in the network and using them like a corpus for word embedding. Two examples of such models are explained in more detail in the following subsections. Moreover, graph neural network (GNN) methods have been actively studied recently and have shown great predictive performances. GNN architectures incorporate message passing between nodes, and node embeddings are generated in a form of aggregating information from neighbors. Several designs of how exactly the messages are passed and aggregated have been proposed, including Graph Convolutional Networks \citep{welling2016semi} and GraphSAGE \citep{hamilton2017inductive}. Though GNNs are capable of leveraging structures of networks well, \citet{liu2023accurately} explain that GNNs typically require node attributes and labels, meaning the resulting embeddings are tied to the quality of those information. The authors also empirically show that given only limited amount of labels, GNNs tend to perform suboptimally: in such cases, random walk based methods, which do not require anything more than the network alone (set of nodes and edges), work better and so these methods have a unique place in node embedding, apart from GNNs. In this paper, we focus on random walk based node embedding methods, especially in the viewpoint of generating embeddings with homophily and edge weights considered.

\subsection{node2vec}

The node2vec method is motivated by the idea of a word embedding technique called word2vec developed by \citet{mikolov2013efficient}. Given a large language corpus, the word2vec Skip-gram model densely embeds each word $w$ that appears in the corpus as it gets trained to predict the co-appearing nearby words using the embedding vector of $w$. Usually, the actual computation of the entire loss is too expensive, and thus the negative sampling trick is used \citep{mikolov2013distributed}, and this is the so-called ``SGNS'' (Skip-gram with negative sampling). Ideally, after the word2vec training procedure, words with similar meaning have close embedding vectors. \citet{grover2016node2vec} proposes node2vec, where each node is treated just like each word, and the random walk sequences are utilized as if they are the sentences in the language corpus for word2vec training. Once the random walks are completed, the remaining training mechanism (model architecture, loss, negative sampling trick, etc) is the same as in word2vec SGNS. As an example, suppose the walk length is set as 30, and the window size is set as 4. Then each \textit{walk} with length 30 is splitted into \textit{subsequences} of length 4. Then, for each subsequence, the initial node pairs up with all the remaining nodes to form a positive example. That is: a walk $[ v_{1}, v_{2}, \cdots, v_{30} ]$ produces positive examples as follows:
\begin{itemize}
  \item $[v_{1}, v_{2}, v_{3}, v_{4}] \rightarrow (v_{1}, v_{2}), (v_{1}, v_{3}), (v_{1}, v_{4})$
  \item $[v_{2}, v_{3}, v_{4}, v_{5}] \rightarrow (v_{2}, v_{3}), (v_{2}, v_{4}), (v_{2}, v_{5})$
  \item $\cdots$
  \item $[v_{27}, v_{28}, v_{29}, v_{30}] \rightarrow (v_{27}, v_{28}), (v_{27}, v_{29}), (v_{27}, v_{30})$
\end{itemize}
For each positive example, negative examples are randomly created, where the number of negatives per positive is also a setting choice.

More specifically, node2vec performs a second-order random walk with two hyperparameters $p$ and $q$ that choose a balance between homophily and structure, which can also be thought as walking globally or locally. Throughout the paper we assume an undirected graph $G = (V, E)$ with non-negative edge weights unless specified otherwise. Let $N(v)$ denote the set of neighbors for any node $v$, and let $w(u,v)$ denote the edge weight for any pair $(u,v)$ of nodes with an edge. Consider performing a random walk in $G$, and let $t$ be the previous node and $v$ be the current node. 
For each $x \in N(v)$, the (unnormalized) probability of choosing $x$ to visit next is: $\pi(x | v, t) = \alpha_{p,q}(t,x) \cdot w(v,x)$, where the bias factor $\alpha_{p,q}(t,x)$ is:

\[
    \alpha_{p,q}(t,x) =
\begin{cases}
    1/p & \text{if } t=x \\
    1 & \text{if } (t,x) \in E \\
    1/q & \text{if } (t,x) \notin E
\end{cases}
\]

\subsection{node2vec+}

The node2vec method can work generally with any (un)directed, (un)weighted network. However, \citet{liu2023accurately} points out that it does not consider the edge weight between the previous node and the potential next node during a random walk and further proposes node2vec+. This method extends node2vec in a way that accounts for weights in walk biases but reduces back to node2vec for unweighted graphs or unbiased walks.

The difference from node2vec is the bias factor $\alpha_{p,q}(t,x)$ formula when choosing the next node to visit during a random walk. In node2vec, for a neighbor $x$ of the current node $v$ that is also a neighbor of the previous node $t$, the bias factor is just $1$. The main question tackled is ``what if $t$ and $x$ have an edge but with a very weak weight?'' That is, $t$ and $x$ are \textit{barely} connected. With this motivation, the notions of `loose' and `tight' edges are introduced. \citet{liu2023accurately} states: an edge $(u,v)$ is defined to be `loose' if $w(u,v) < \tilde{d}(u)$ where $\tilde{d}(u) = \frac{\sum_{x \in N(u)} w(u, x)}{|N(u)|}$ denotes the average weight of all edges of $u$. Naturally, for an undirected edge $(u,v)$, we can say $(u,v)$ is loose if $w(u,v) < \text{max}(\tilde{d}(u), \tilde{d}(v))$ since it makes no sense to have `edge $(u,v)$ is loose but edge $(v,u)$ is not'. All edges that are not loose are defined to be `tight'. Further, for ease of notation, any node pair without an edge is also said to be `loose'. The bias factor formula for node2vec+ can then be written as: 

\[
    \alpha_{p,q}(t,x) =
\begin{cases}
    1/p & \text{if } t=x \\
    1 & \text{if } (t,x) \text{ tight} \\
    \text{min}(1, 1/q) & \text{if } 
    \begin{aligned}[t]
       & (t,x) \text{ loose,} \\
       & (v,x) \text{ loose}
       \end{aligned} \\
    1/q + (1 - 1/q) (\frac{w(x,t)}{\text{max}(\tilde{d}(x), \tilde{d}(t))}) & \text{if } \begin{aligned}[t]
       & (t,x) \text{ loose,} \\
       & (v,x) \text{ tight}
       \end{aligned} 
\end{cases}
\]

It is demonstrated that the node2vec+ method performs comparatively or better than node2vec in several datasets.

\section{Proposed Method} 

\subsection{Motivation: When do two words get similar word2vec embeddings?} \label{motivation}

Consider two words that are similar in meaning, say, ``baseball'' and ``basketball''. With a good-quality corpus that contains many sentences, when we train a word2vec SGNS, those two words will obtain embeddings that are close to each other. This is because they often appear in similar \textit{context}, and specifically for word2vec, that \textit{context} is represented as the context window i.e. words that appear within a few positions before and after the word. For example, we won't be surprised to observe the following phrases together in the corpus: ``favorite baseball player'' and ``favorite basketball player''. Another example is: ``the new baseball stadium'' and ``the new basketball stadium''. Eventually, the two words will appear in many training examples with common neighbor words i.e. the other input word in SGNS (in our example, words like ``player'' and ``stadium'') and so at the end the two will have close embeddings. That is: \textbf{for two words to get similar word2vec embeddings, the corpus needs to contain sentences where the two words share similar context windows}. This principle motivated our proposed method which is explained in the following subsection.

\subsection{ARGEW: Augmentation of Random walks by Graph Edge Weights}

\subsubsection{Background}

As stated in the ``Introduction'' section, our goal is to obtain embeddings such that two nodes with stronger edge weights tend to have closer embeddings. That is: a node pair without an edge should have very far embeddings, a pair with a weak edge should have little closer embeddings, and the embeddings for a pair with a strong edge should be very close. Such embeddings are especially meaningful for homophilous graphs where node similarity is proportional to the strength of connections, as we can directly aim other tasks like node clustering or classification.

In node2vec, once the corpus, a set of random walks performed in the graph, is established, the training mechanism afterwards is identical to SGNS. Hence, the principle mentioned in section \ref{motivation} can be naturally stated in node2vec setting too: \textbf{two nodes get similar node2vec embeddings if they share similar sets of nodes co-occuring throughout the random walks}.

With this idea, we propose ARGEW (Augmentation of Random walks by Graph Edge Weights), a novel augmentation method for random walks with the goal of making nodes with strong edges end up with similar embeddings. We perform augmentation on random walk data (i.e. derive new subsequences) such that at the end, for two nodes with strong edges, there are many walk sequences such that the two are surrounded by the same set of neighbor nodes. We are, in some sense, enforcing node pairs with large edge weights to satisfy the condition for two nodes obtaining similar node2vec embeddings. Further, we make sure that the enforcement solely relies on the graph structure, specifically neighborhoods and edge weights, so that the augmentation is perfectly plausible. 

\subsubsection{The ARGEW algorithm}

Given a random walk subsequence, we loop through each node, replace it with its neighbor with the largest weight, and add the new subsequence to the training data. To ensure that the augmentation is completely realistic, we only consider neighbors that are also neighbors of the preceding and following nodes. For example, for a subsequence $[v_{2}, v_{3}, v_{4}, v_{5}]$, when we are at the second spot i.e. $v_{3}$, we only consider the nodes $x \in N(v_{3}) \cap N(v_{2}) \cap N(v_{4})$ and pick $v_{3}'$ with the largest edge weight with $v_{3}$. The obvious reason is to guarantee that the new derived subsequence $[v_{2}, v_{3}', v_{4}, v_{5}]$ is perfectly plausible i.e. not surprising to show up in a real random walk. If we are at the first (last) position in the subsequence, then we only consider neighbors that are also neighbors of the next (previous) node since the previous (next) node is undefined. Algorithm~\ref{algo:argew} shows the pseudocode for ARGEW.

\paragraph{Deciding augmentation counts via exponent}

A critical question is how much we should increase the new subsequences in the training set.
Let $w_{v_3, v_3'}$ represent the highest weight among the potential neighbors in the example from the preceding paragraph. Based on the idea that two words (nodes) with many common co-occurring items will end up with similar embeddings, we may anticipate that the more augmentation we do in ARGEW, the similarity between the embeddings of $v_{3}$ and $v_{3}'$ will increase. Therefore, the higher value of $w_{v_3, v_3'}$, we will add more of the new subsequence i.e. $[v_{2}, v_{3}', v_{4}, v_{5}]$ to the corpus (repetitively add the same subsequence). We implement this idea as follows. First, we apply min-max scaling to rescale each unique edge weight in the network to a value between $[low, high]$, where $low$ and $high$ are hyperparameters. This is done by using the following formula for each weight value $x$: 
\vskip -10pt
\begin{equation}
x' = \big((x - min) / (max - min)\big) \cdot \big(high - low\big) + low
\end{equation}
where $min$ and $max$ are the smallest and largest edge weight values in the entire graph, respectively. For example, if a network's unique weights are $0.15, 0.35, 0.6, 0.8, 0.85, 0.9$ and we choose $low$ = $1$, $high$ = $7$, then the rescaled weights are $1, 2.6, 4.6, 6.2, 6.6, 7$. Then, at each augmentation step, the original subsequence is added once, followed by the addition of the new subsequence $2^{r(w_{v_3, v_3'})}$ times, where $r(w_{v_3, v_3'})$ is the rescaled value of the weight $w_{v_3, v_3'}$. We use the exponent in order to widen the difference between the amount of augmentation when $w_{v_3, v_3'}$ is small and when $w_{v_3, v_3'}$ is large. In the later experiments section, we show that adjusting the rescale range ($low$ and $high$) indeed affects the difference between the embeddings' similarity for small edge weights and large edge weights.

\paragraph{Skipping small edge weights}

When $w_{v_3, v_3'}$ is tiny, adding the derived subsequence to the corpus is undesirable, which is similar to why we use the exponent. Though the exponent suppresses the augmentation count for small $w_{v_3, v_3'}$, often the $w_{v_3, v_3'}$ will be small because it is extremely likely that the network has significantly more small-weight edges than large-weight edges. Since we want more augmentation for cases with greater $w_{v_3, v_3'}$, adding up a lot of new subsequences with small $w_{v_3, v_3'}$ is not useful for us. Hence we only perform augmentation when $w_{v_3, v_3'}$ is strictly greater than the median of the (non-unique) edge weights in the whole network.

\begin{algorithm}[ht]
\caption{The ARGEW algorithm}\label{alg:cap}
\label{algo:argew}
\begin{algorithmic}
\Function{argew}{graph $G$, $low$, $high$}
\State $w \gets \text{edge\_weight}(G)$
\State $r \gets \text{rescale\_weight}(w, low, high)$ 
\State $w_{med} \gets \text{median}(w)$
\State $subsequences \gets \text{do\_random\_walk}(G)$ 
\For{$subseq$ in $subsequences$} 
\State add $subseq$ to training corpus 
\For{$v$ in $subseq$} 
\State $v', w_{v,v'} \gets \text{find\_substitute}(G, subseq, w, v)$ 
\If{$w_{med} < w_{v,v'}$}
\State \# \texttt{add subseq once more}
\State add $subseq$ to training corpus 
\State $subseq' \gets replace(subseq, v, v')$
\For{$\_$ in range($0, 2^{r(v,v')}$)} 
\State add $subseq'$ to training corpus 
\EndFor
\EndIf
\EndFor
\EndFor
\EndFunction
\\
\State \# \texttt{find $v' \in N(v) \cap N(\text{prev}) \cap N(\text{next})$ with largest $w(v,v')$}
\Function{find\_substitute}{$G$, $subseq$, $w$, $v$}
\State $v_{prev} \gets \text{previous}(subseq, v)$
\State $v_{next} \gets \text{next}(subseq, v)$
\State $candidates \gets \text{get\_neighbors}(G, v)$
\If{$v_{prev} \text{ is not } null$}
\State $nb_{prev} \gets \text{get\_neighbors}(G, v_{prev})$
\State $candidates \gets \text{intersect}(candidates, nb_{prev})$
\EndIf
\If{$v_{next} \text{ is not } null$}
\State $nb_{next} \gets \text{get\_neighbors}(G, v_{next})$
\State $candidates \gets \text{intersect}(candidates, nb_{next})$
\EndIf

\State $v' \gets null$
\State $w_{v,v'} \gets 0$
\For{$c$ in $candiates$}
\State $w_{v,c} \gets w(v, c)$
\If{$w_{v,v'} < w_{v,c}$}
\State $v' \gets c$
\State $w_{v,v'} \gets w_{v,c}$
\EndIf
\EndFor
\State return $v', w_{v,v'}$
\EndFunction
\end{algorithmic}
\end{algorithm}

\section{Experiments}

\subsection{Datasets}
For evaluating the embedding results, we mainly focus on the similarity distributions of the embeddings because: 1) that is what the desired properties discussed in the ``Introduction'' section explicitly state, and 2) random-walk based embedding methods are especially valuable when we lack good-quality information about node labels and/or attributes. However, we think checking node classification performance as a secondary criterion is also important, as embeddings should ultimately be used for downstream tasks. Hence, to demonstrate how the node embedding results differ based on whether or not the ARGEW method is used, we use three weighted homophilous networks each with node labels.

\begin{figure}[ht]
\centering
\begin{tabular}{c @{\hspace{0.5\tabcolsep}} c} \\
\includegraphics[width=3.9cm]{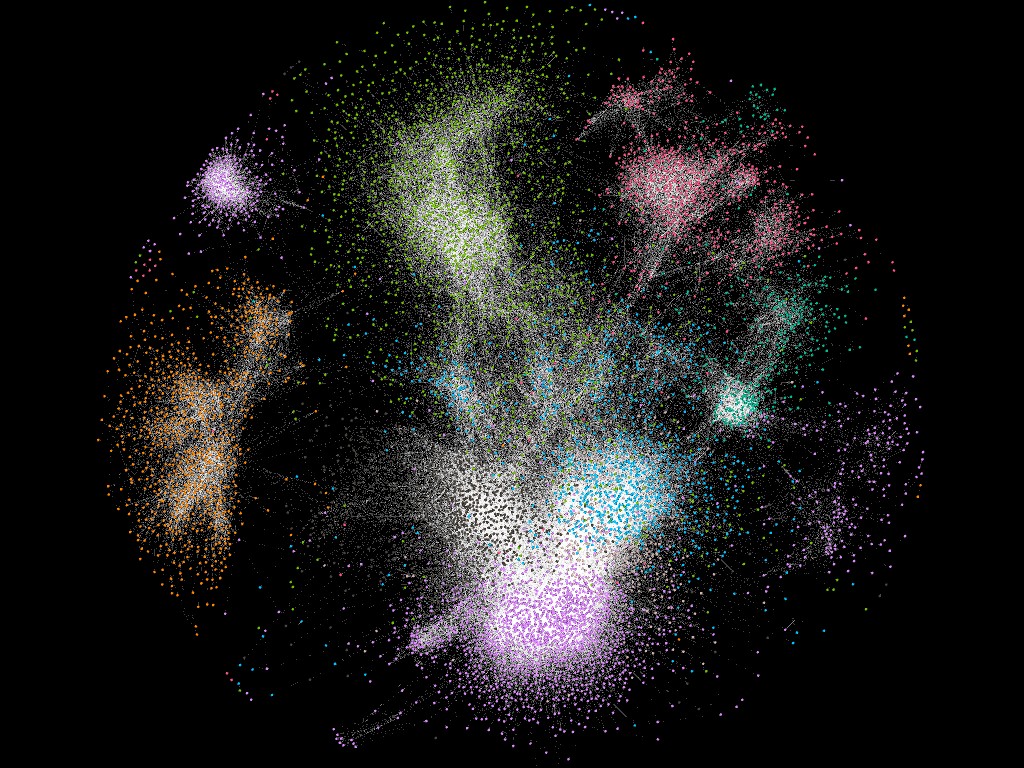} &
\includegraphics[width=3.9cm]{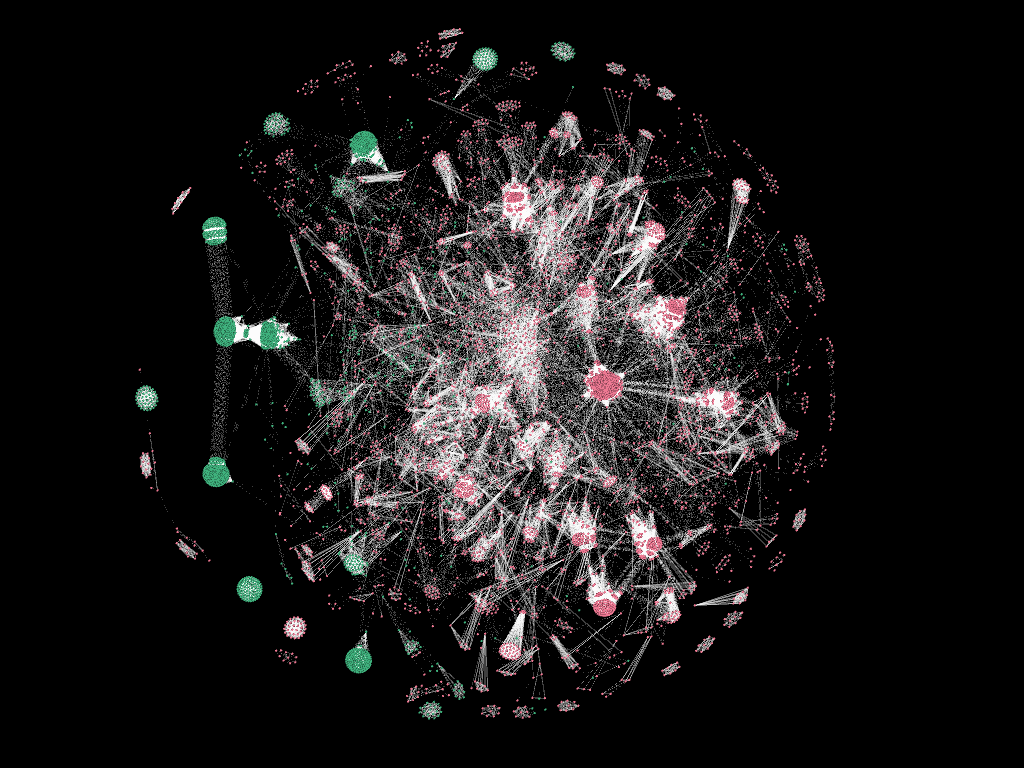}
\\
(a) Amazon Photo network & (b) LineageW network \\
\end{tabular}
\caption{Network visualization (node colors are labels)}
\label{network_gephi}
\end{figure}

Two benchmark datasets, the Cora network \citep{sen2008collective} and the Amazon Photo network \citep{shchur2018pitfalls}, are used. Both are classified to have high homophily in \citet{platonov2022characterizing}. Though the two graphs originally lack edge weights, we designed weights based on the data's context for our experiments. Moreover, we use one network based on game user logs to show the method's usefulness in industry service data. The game used in our experiments is \textit{Lineage W}, which is an MMORPG released on November 2021 by NCSOFT and is ranked \#1 in the ``Top Grossing Games Worldwide for H1 2022'' (Google Play Revenue) chart reported by Sensor Tower \citep{chan2022global}. Figure~\ref{network_gephi} shows the Amazon Photo and LineageW networks visualized in the \textit{Gephi} software \citep{bastian2009gephi}, in which homophily is clearly shown. Visualization of the Cora network indicating its homophily is avaiable in previous works such as \citet{ringsquandl2021power}.

\subsubsection{Cora network} 
Each node is a scientific publication labeled as the paper topic (among 7 classes), and edges indicate citations. We use the ``Planetoid'' version provided in the \textit{PyG} (\textit{PyTorch Geometric}) python package \citep{fey2019fast}, which has 2,708 nodes and 5,278 edges. Each node's attribute is a 1,433-dimensional vector indicating whether the corresponding word appears in the publication (based on a dictionary of 1,433 words). For each edge, we defined the weight as the number of dimensions, out of 1,433, such that the two attribute vectors' values are equal. That is: how many words do the two papers either both use or both not use.

\subsubsection{Amazon Photo network}
Nodes represent photo-related products in Amazon, and edges mean the two products were bought together often. For each of the 7,535 nodes, whose label is the specific product category (among 8 classes), the attribute is its 745-dimensional bag-of-words (BoW) vector of the product's review texts. We defined edge weights as the cosine similarity between the two BoW vectors. Among the 119,082 edges, 2 have zero cosine similarity, which we remove, ending up with 119,080 edges.

\subsubsection{LineageW network}
As explained by \citet{lee2016you, lee2018no}, in a typical MMORPG where a user develops a character in the virtual world (i.e. a particular \textit{game server}), users naturally combat and socialize with others, and that process establishes the in-game economy which is connected to the real economy in several aspects. Note that a \textit{person} can have multiple \textit{accounts}, and one account can have many \textit{characters}. There are many types of social interactions, or relationships, among users, and we constructed a network representing such interactions somewhat comprehensively. We chose one date and one game server of \textit{Lineage W}. The 6,792 nodes are accounts, and the 338,809 edges indicate the two accounts had at least one of the seven types of interaction that we defined as follows:

\begin{enumerate}
    \item In-game trade: did an in-game trade suspected to be solely for game-money delivery, which is typical in gold-farming groups as explained by \citet{lee2018no}
    \item Co-hunting: participated in a group of 3 or more characters that together killed at least 100 monsters
    \item Similar character names: have characters with name similarity $> 0.6$ where similarity is a score of $1-$ normalized Levenshtein distance
    \item Customer Id: have the same (hashed) customer id
    \item Device Id: have the same (hashed) device id
    \item IP: entered the game with the same IP address
    \item IP b class: entered the game with IP addresses with the same b class (i.e. the first two of the four numbers in the IP)
\end{enumerate}
Each edge's weight represents how many, out of seven, types of interactions the two nodes had.


\paragraph{Node labels: non-client bot} For the node classification task, we label each node on whether the account is suspected as a non-client bot. 
We suspect a user to be a non-client bot if we observe plenty of records about their play but the corresponding data that are supposed to be sent from the game client are missing nearly entirely. This means the user played the game without a normal client program but via only using network packet transactions with the server, which is intentional abusing. Note that bots that work together as a group typically form character names with a common random meaningless string followed by a number for identification, as explained by \citet{kang2017char}, which is why we include similar character names as an interaction type for the network data.

\subsection{Experimental setup}
For each of the three networks, we compare the results of node2vec with and without ARGEW augmentation.

\paragraph{Training node2vec without ARGEW} We use the following hyperparameters: dimension of embedding = $128$, length of each walk = $80$, context size = $10$, number of walks per node = $10$, number of negative samples per positive = $1$. We experiment all $p, q \in \{ 0.25, 1, 4 \}$. We use a learning rate of $0.01$ and train for 10 epochs with early stopping if the loss is not smaller than the previous epoch's.

\paragraph{Training node2vec with ARGEW} Since ARGEW derives additional subsequences, we can, and should (due to memory limit), reduce the amount of walks compared to vanilla node2vec. Hence, we change the number of walks per node to $1$ and use a smaller batch size. The other hyperparameters remain the same as in without ARGEW. For weight rescaling, we use $low$ = $1$, $high$ = $9$.

\paragraph{Node label classification} We train a one vs rest logistic regression classifier with L2 regularization where the node embedding vector is used as (the only) features. We perform 10 times of stratified sampling that splits the data to 50\% training and 50\% testing. The micro-averaged F1 score (of the 10 test scores) and the macro-averaged F1 score (of the 10 test scores) are reported.

\subsection{Results}

\subsubsection{Embedding cosine similarities}
\begin{figure}[ht]
\centering
\begin{tabular}{c @{\hspace{0.5\tabcolsep}} c} \\
\includegraphics[width=7cm]{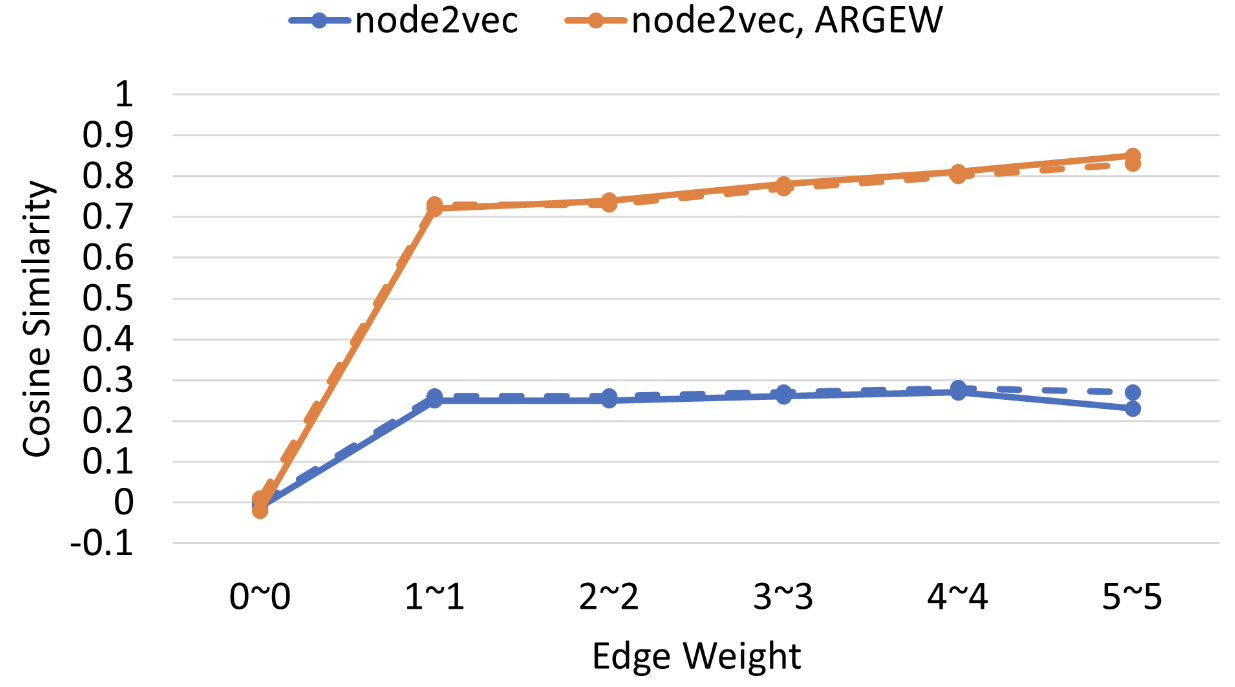} \\
(a) LineageW network ($p=1, q=1$) \\ \\
\includegraphics[width=7cm]{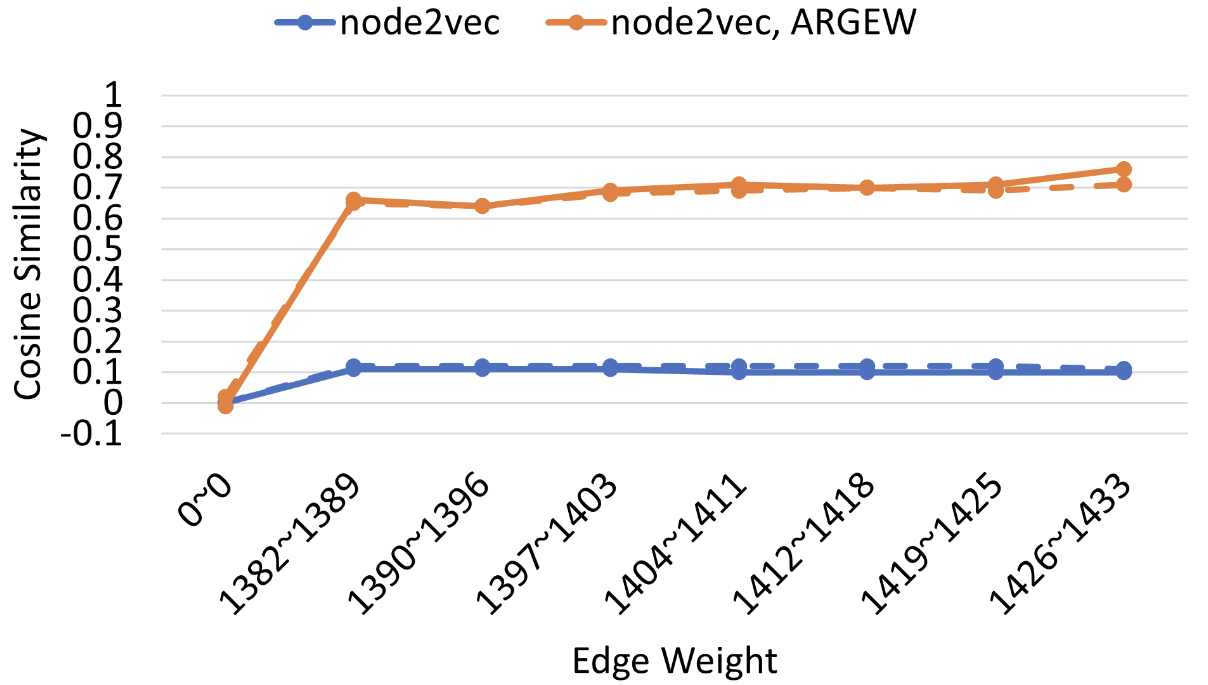} \\ 
(b) Cora network ($p=0.25, q=0.25$) \\ \\
\includegraphics[width=7cm]{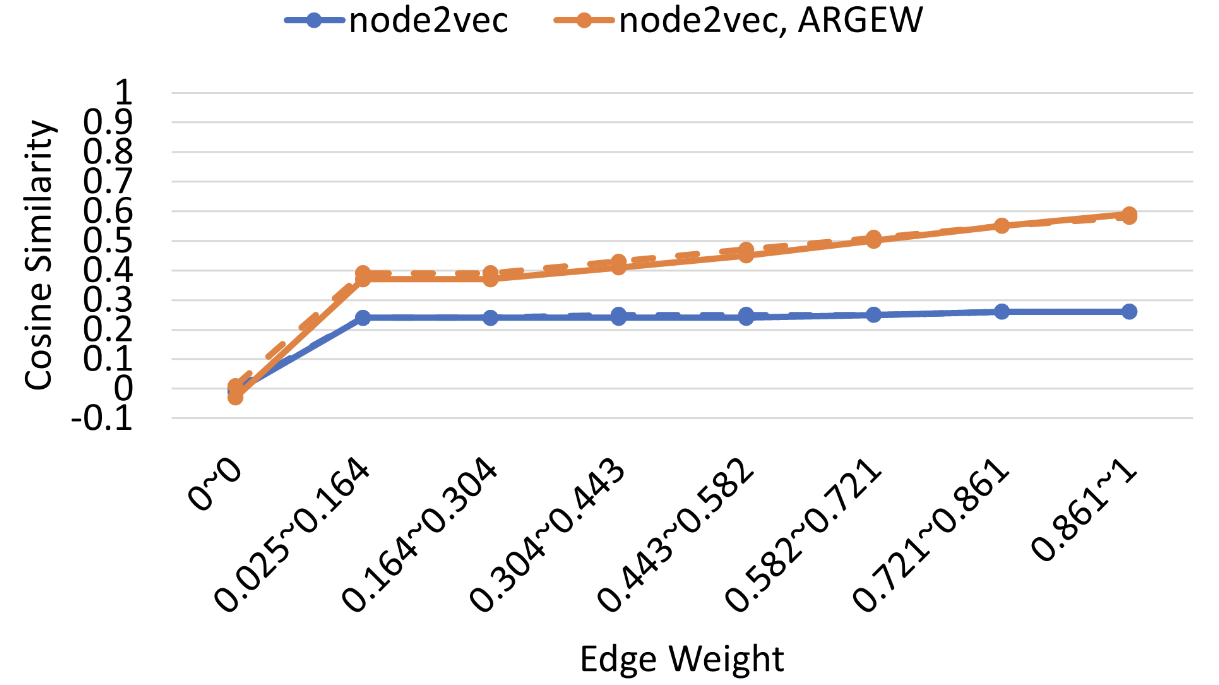} \\
(c) Amazon Photo network ($p=1, q=1$) \\
\end{tabular}
\caption{Cosine similarity results (solid: median, dashed: mean)}
\label{simdist2}
\end{figure}

We equally split the edge weights to bins and visualize the median and the mean of the embedding cosine similarities for node pairs within that weight range. Among the $p, q \in \{ 0.25, 1, 4 \}$ settings, we plot for the setting with the highest classification micro-F1 score. Figure~\ref{simdist2} compares the results with and without ARGEW: the x-axis specifies the range of edge weights, and the y-axis is the cosine similarity value for all edges with the corresponding weights. Note that the leftmost bin (i.e. weight = $0$) represents all node pairs that do not have an edge. We can observe that results with ARGEW have a clearer pattern that: pairs with edges have larger similarities than pairs without edges, and furthermore, weak edges have smaller similarities compared to strong edges, in terms of both the median and the mean. This is common across all datasets, which demonstrates that ARGEW can successfully reflect edge weights in the node embeddings.

\begin{figure}[ht]
\centering
\begin{tabular}{c @{\hspace{0.5\tabcolsep}} c} \\
\includegraphics[width=7cm]{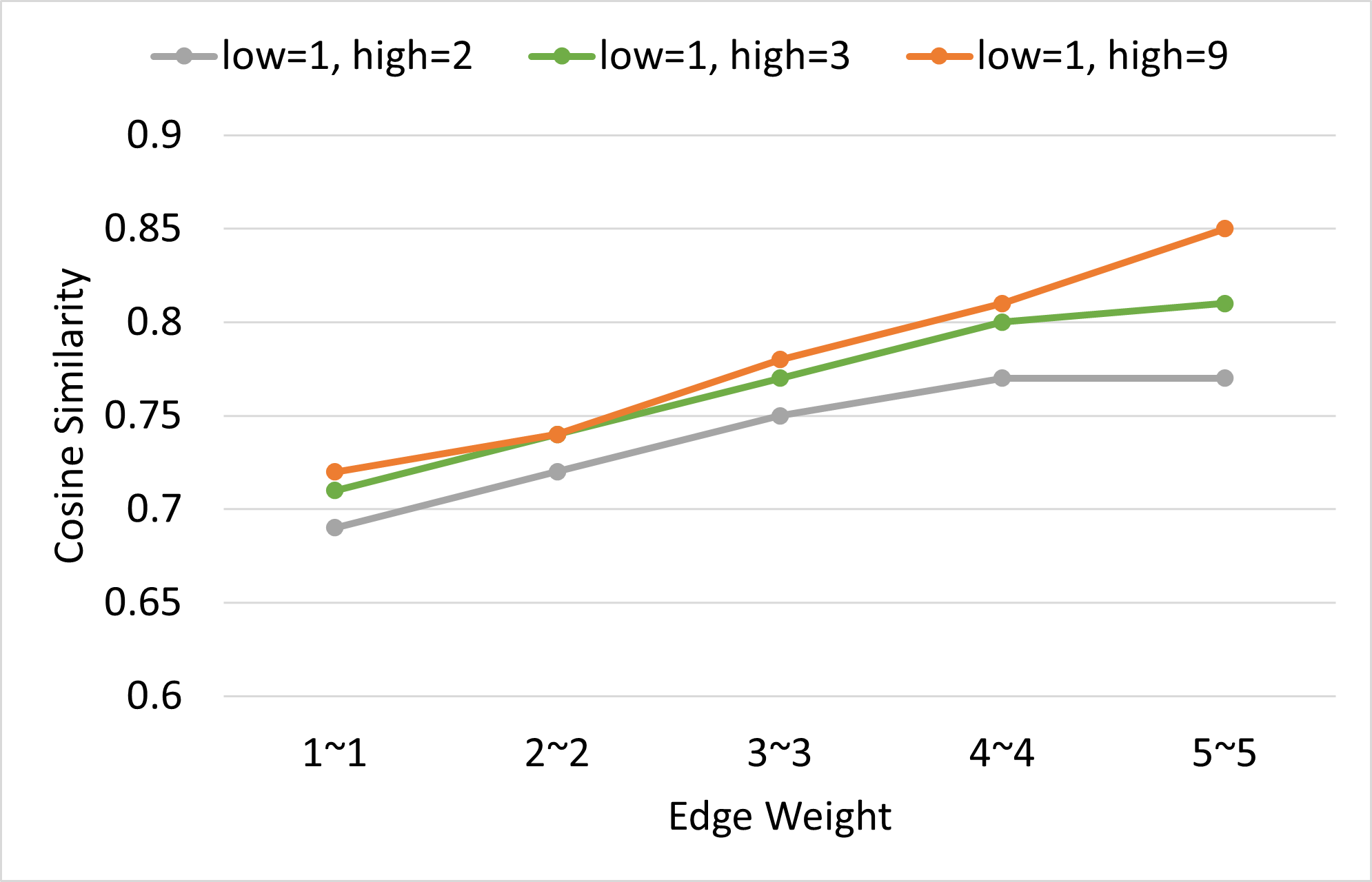} \\
\end{tabular}
\caption{Embedding similarity medians for different rescale range settings (LineageW network; $p=1, q=1$)}
\label{rescalerange}
\end{figure}


\begin{figure}[ht]
\centering
\begin{tabular}{c @{\hspace{0.5\tabcolsep}} c} \\
\includegraphics[width=7cm]{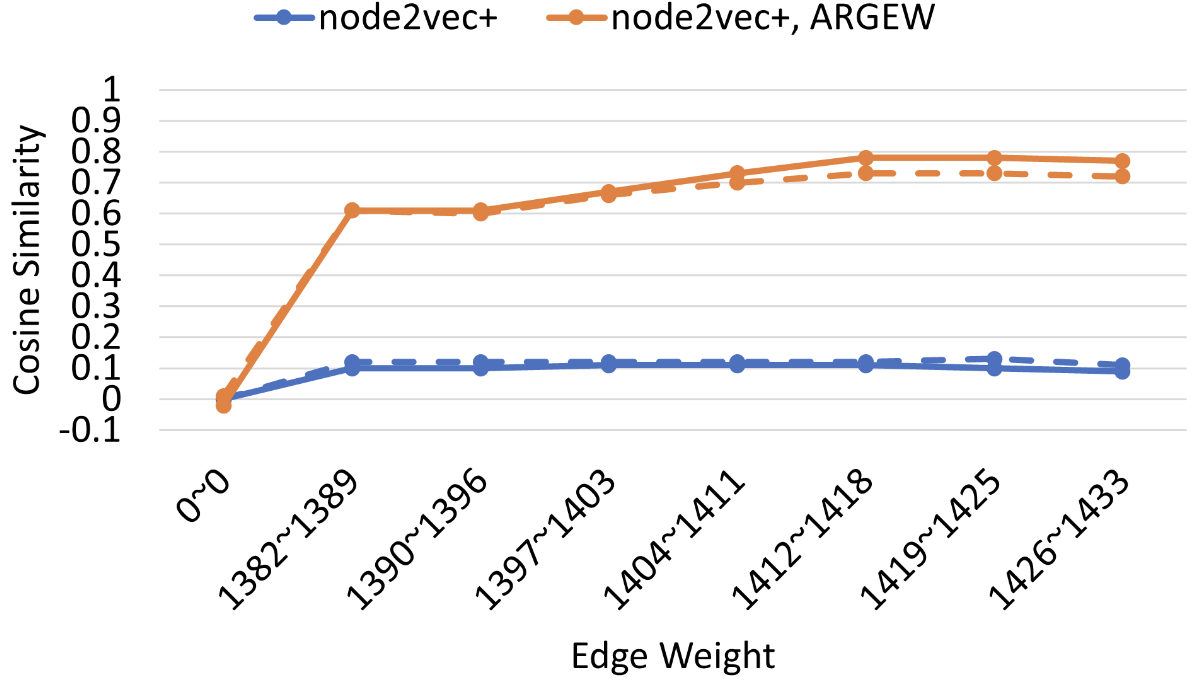} \\
\end{tabular}
\caption{Cosine similarity results (solid: median, dashed: mean) for node2vec+ (Cora network; $p=4, q=0.25$)}
\label{coranode2vecplus}
\end{figure}


\paragraph{Impact of the rescale range} 
In Figure~\ref{simdist2}, for all results that use ARGEW, we set the rescale range as $low$ = $1$, $high$ = $9$. As explained in the ``Proposed Method'' section, the purpose of ARGEW's rescaling is to widen the gap between the amount of additions for small-weight edges and that for large-weight edges. Figure~\ref{rescalerange} compares the similarity median results when $high$ = $2$, when $high$ = $3$, and when $high$ = $9$ (all with $low$ = $1$) for node2vec with ARGEW on the LineageW network. Note that we excluded the non-edge pairs (i.e. weight = $0$) from the plot since the median is approximately $0$ for all cases. Compared to $high$ = $2$, the increasing trend is a little clearer in $high$ = $3$ and much more explicit in $high$ = $9$. This indicates that wider rescale ranges can widen the gap between the embedding similarity for small-weight edges and that for large-weight edges. Note that the larger the $high$ value is, the more the GPU resource is used, and hence we set the range by considering our GPU memory.

\paragraph{ARGEW can work with any random walk strategy} 
ARGEW is independent of the walk sampling strategy and operates on top of the already-prepared random walks. As an example, we perform node2vec+, instead of node2vec, on the Cora network and show the embedding similarity results with and without ARGEW in Figure~\ref{coranode2vecplus}. Again we display the $p, q \in \{ 0.25, 1, 4 \}$ setting that yields the highest classification micro-F1 score. The desired increasing trend is similarly shown as in Figure~\ref{simdist2}(b), indicating that ARGEW successfully works with not just node2vec but any walk strategy.

\subsubsection{Node classification performance}

Though our primary evaluation criterion for ARGEW is the embedding similarity results, we also check its performance gain in node classification as good embeddings should ultimately be used in downstream tasks.
Tables~\ref{microf1} and~\ref{macrof1} show the micro and macro averaged F1 scores for the node classification tasks for the $p, q \in \{ 0.25, 1, 4 \}$ setting where ARGEW had the highest micro F1 score for each dataset, just like in Figure~\ref{simdist2}. We can see that for all three networks, the performance when ARGEW is used is clearly better than pure node2vec. Nodes connected with large edge weights gained closer embeddings, and this naturally led to better classification as nodes with the same label tend to be connected strongly by homophily.

\begin{table}[h]
\caption{Node Classification Micro-averaged F1 Scores} \label{microf1}
\begin{center}
\begin{tabular}{cccc}
\hline
Model         
& \begin{tabular}[c]{@{}c@{}} LineageW \end{tabular} 
& \begin{tabular}[c]{@{}c@{}} Cora \end{tabular}
& \begin{tabular}[c]{@{}c@{}} Amazon \end{tabular} \\
\hline \hline 
node2vec             & $0.918$ & $0.533$ & $0.825$ \\
\textbf{node2vec, ARGEW}             & $\mathbf{0.960}$ & $\mathbf{0.752}$ & $\mathbf{0.864}$ \\
\hline
GCN (supervised)            & $0.978$ & $0.776$ & $0.898$ \\
\hline
\end{tabular}
\end{center}
\end{table}

\begin{table}[h]
\caption{Node Classification Macro-averaged F1 Scores} \label{macrof1}
\begin{center}
\begin{tabular}{cccc}
\hline
Model         
& \begin{tabular}[c]{@{}c@{}} LineageW \end{tabular} 
& \begin{tabular}[c]{@{}c@{}} Cora \end{tabular}
& \begin{tabular}[c]{@{}c@{}} Amazon \end{tabular} \\
\hline \hline 
node2vec             & $0.895$ & $0.487$ & $0.747$ \\
\textbf{node2vec, ARGEW}             & $\mathbf{0.948}$ & $\mathbf{0.733}$ & $\mathbf{0.832}$ \\
\hline
GCN (supervised)            & $0.972$ & $0.755$ & $0.868$ \\
\hline
\end{tabular}
\end{center}
\end{table}


\paragraph{Comparison to GCN} 

For more comparison, we also check the classification performance of a Graph Convolutional Network (GCN) model on each dataset. Edge weights, node attributes and labels are fully provided, and the model is trained as supervised classification. The model has two graph convolution layers, where the unit dimensions are: the node attribute dimension, then 700 hidden units, and finally the number of classes. For the LineageW network, which lacks node attributes, we use one-hot vectors, whose dimension is the number of nodes, as attributes. We follow the same 10-time 50\%-50\% stratified sampling and check the average of the 10 scores just like our node2vec classification experiments. Tables~\ref{microf1} and~\ref{macrof1} show that GCN's F1 scores are only slightly higher than ARGEW. This indicates when we use ARGEW, node2vec embeddings can achieve classification performance similarly high as GCN, \textbf{even without any node attribute and label information during training}.

\subsubsection{Parameter sensitivity}

In \citet{grover2016node2vec}, the original node2vec paper, the authors provide parameter sensitivity experiments, which demonstrate that node2vec is indeed sensitive to parameter settings. For each hyperparameter, the node classification performance is compared across different value settings (all other hyperparameters are fixed as default values). We perform similar experiments, both with and without ARGEW, in order to thoroughly examine the behaviors of our augmentation technique.

\begin{figure}[ht]
\centering
\begin{tabular}{c @{\hspace{0.5\tabcolsep}} c} \\
\includegraphics[width=4cm]{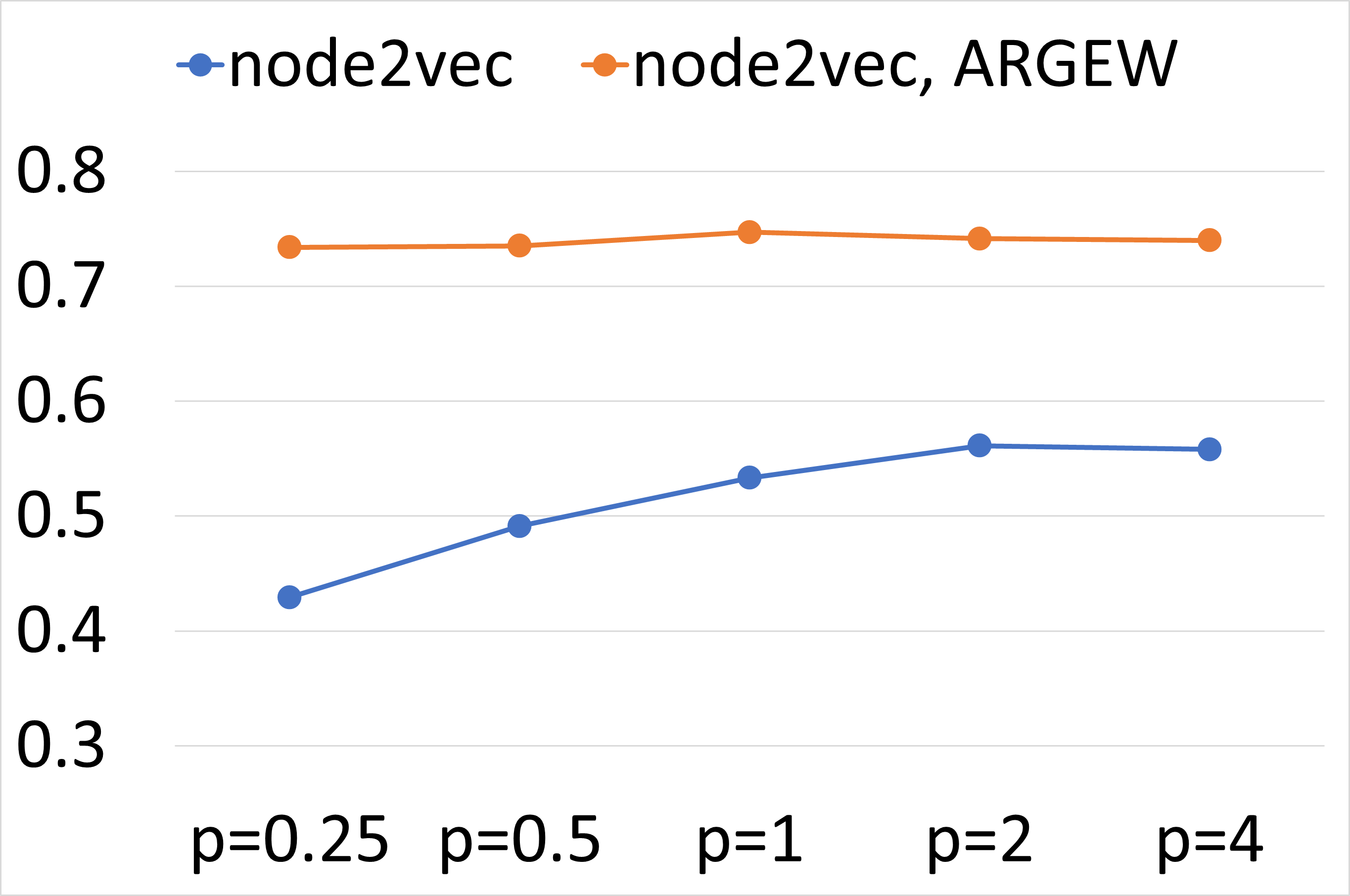} &
\includegraphics[width=4cm]{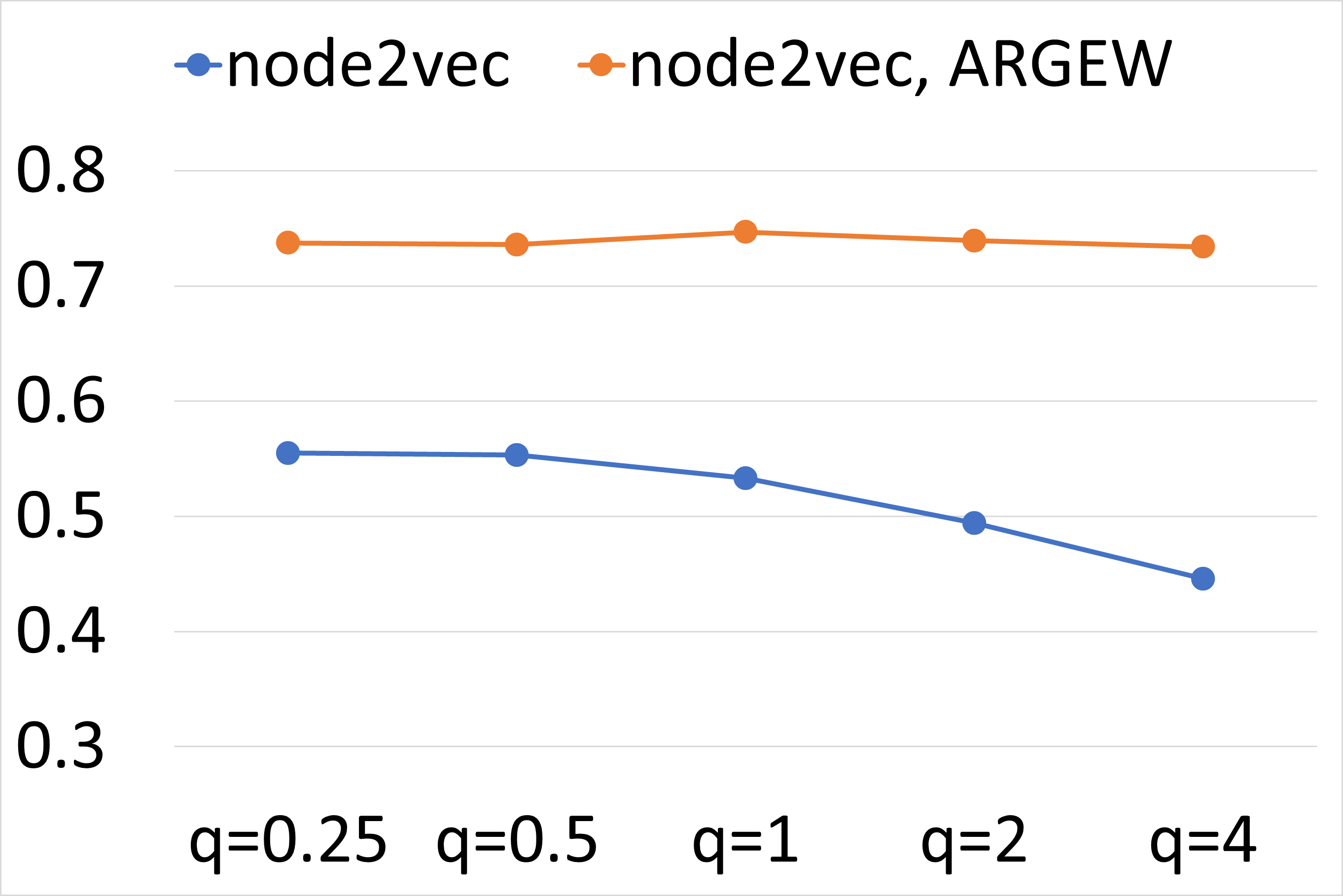} 
\\
(a) $p$ & (b) $q$ \\ \\
\includegraphics[width=4cm]{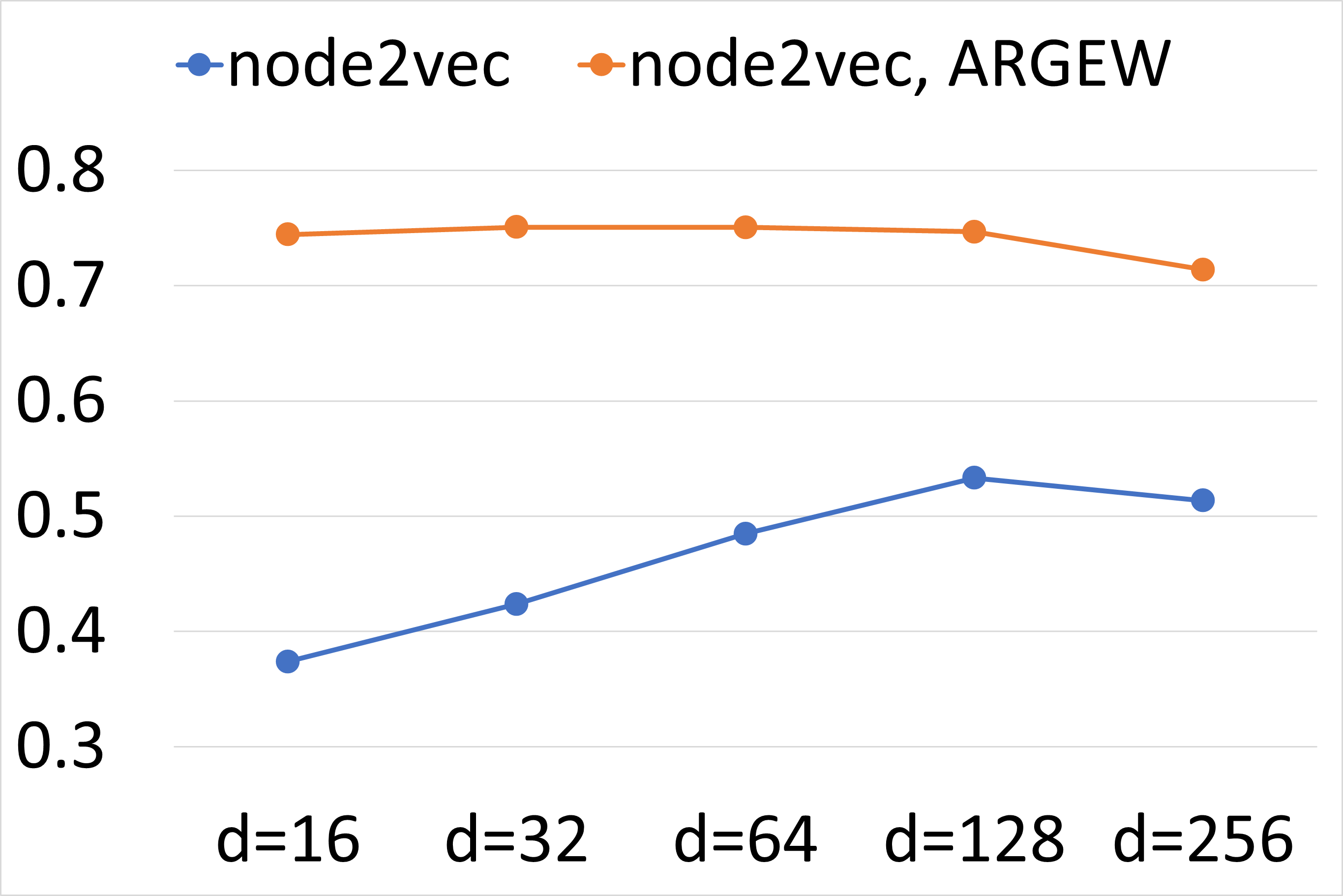} &
\includegraphics[width=4cm]{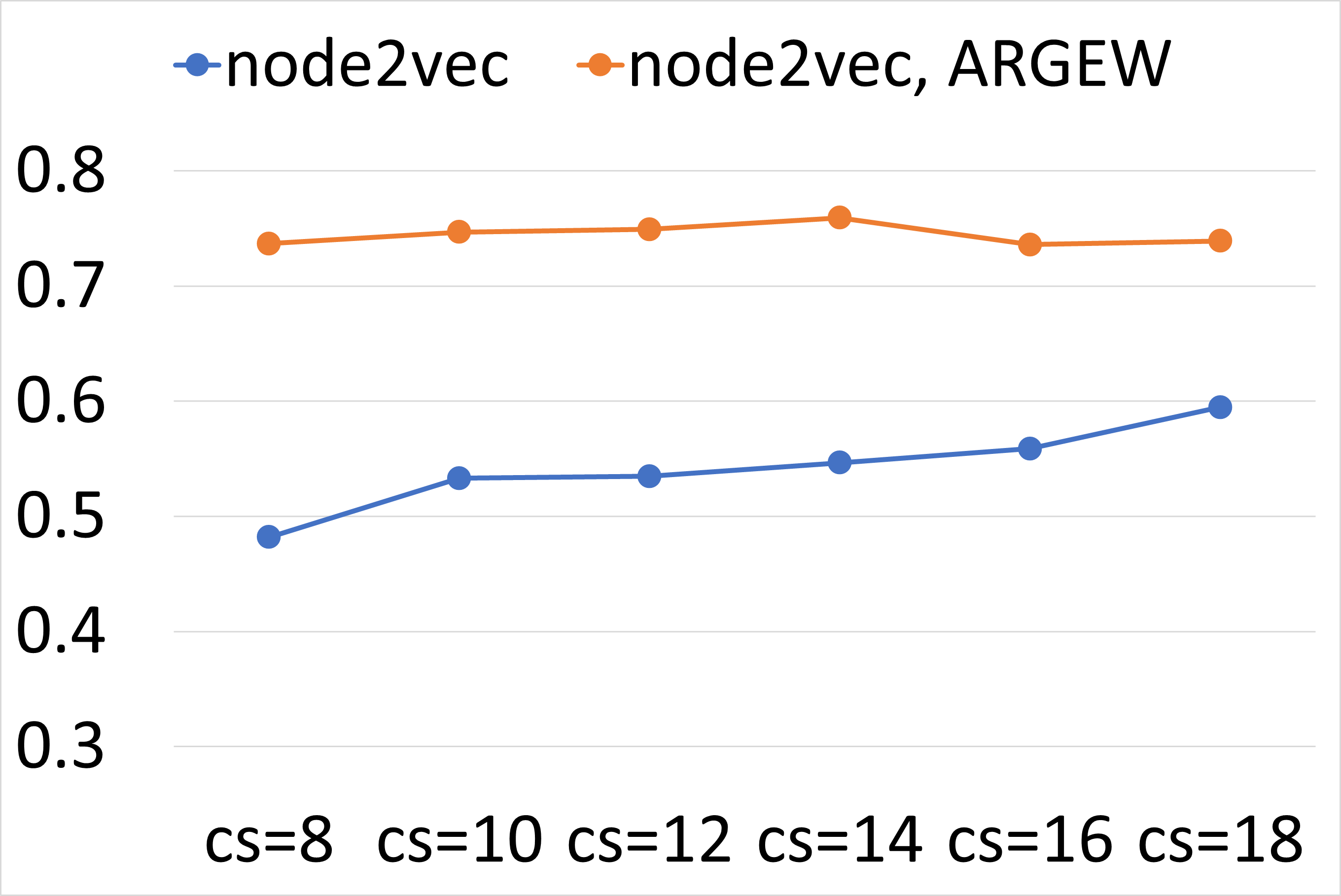}
\\
(c) embedding dimension & (d) context size \\
\end{tabular}
\caption{Parameter sensitivity: Micro-F1 scores (Cora network)}
\label{paramsensit}
\end{figure}

On the Cora network, we train node2vec, both with and without ARGEW, and perform node classification. When ARGEW is used, the performances are not sensitive to any of the hyperparameters: for all settings, the F1 scores are between $0.7$ and $0.8$. On the other hand, for node2vec without ARGEW, not only are the scores all lower than $0.6$, but are also sensitive to hyperparameters, especially $p$, $q$, embedding dimension, and context size, as shown in Figure~\ref{paramsensit}. In other words, ARGEW is non-sensitive to parameter setting with consistently higher performances, compared to vanilla node2vec.

Without ARGEW, node2vec tends to perform worse for smaller $p$, embedding dimension, context size, and larger $q$. This is not surprising according to the explanations in \citet{grover2016node2vec}. Smaller embedding dimension and context size simply mean having less sampling budget, resulting in less informative representations. When $p$ is small and $q$ is large, walks tend to stay locally instead of going outwards, and hence embeddings focus on structural equivalence rather than communities (homophily).

\subsubsection{Walk experiments (+ Why ARGEW really works)}
To understand why ARGEW is not sensitive to hyperparameters, particularly $p$ and $q$ which indeed heavily affects vanilla node2vec, we examine the characteristic of the walks in different settings. 

\begin{figure}[ht]
\centering
\begin{tabular}{c @{\hspace{0.5\tabcolsep}} c} \\
\includegraphics[width=5cm]{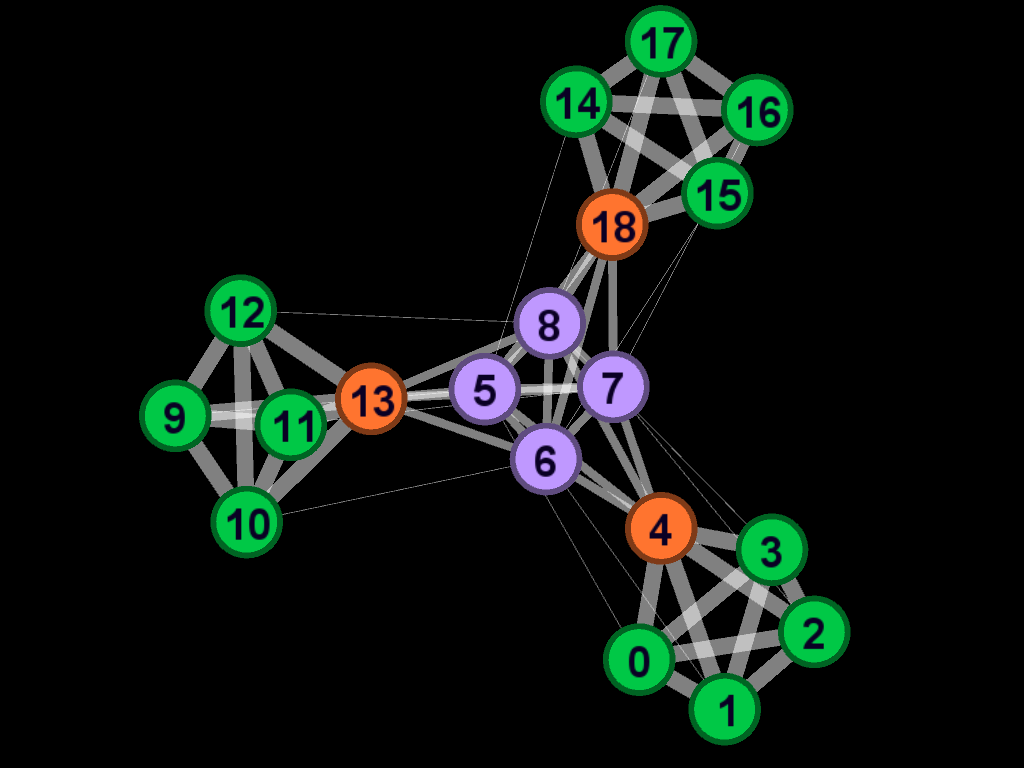} 
\end{tabular}
\caption{A synthetic network with clear structural roles (green: ``internal'', orange: ``bridge'', purple: ``etc'')}
\label{walksamplenetwork}
\end{figure}

We use a simple synthetic undirected network with clear structural roles, as shown in Figure~\ref{walksamplenetwork}. There are three communities, where the orange nodes (\#4, \#13, \#18) are the ``bridge'' node of each community. The green nodes are ``internal'' nodes in their corresponding communities, and the four purple nodes (\#5, \#6, \#7, \#8) are those without a community, which we call ``etc'' nodes. In each community, i.e. the single ``bridge'' node and four ``internal'' nodes, every node pair has an edge of weight = $3$. Further, each ``bridge'' node has an edge of weight = $2$ with all four ``etc'' nodes. Lastly, each ``internal'' node has an edge of weight = $1$ with only one ``etc'' node. In homophily context, we desire each ``bridge'' node to end up with similar embeddings with the ``internal'' nodes in its community. 


For each node $v$, we focus on the nodes that $v$ coappears in walk subsequences and hence go together as a positive pair for SGNS training. More specifically, we categorize the nodes to 7 types based on their roles and communities, as in Table~\ref{typecateg}. For example, type ``c\#4internal'' is the internal nodes in the community whose bridge is node \#4. For each type $T$, we gather the subsequences whose first node has type $T$, and compute the proportion of each type of the coappearing nodes throughout all those subsequences.

\begin{table}[h]
\caption{Node type categorization} \label{typecateg}
\begin{center}
\begin{tabular}{cccc}
\hline
Type         
& Nodes \\
\hline \hline 
c\#4internal             & \#0, \#1, \#2, \#3 \\
c\#4bridge            & \#4  \\
c\#13internal             & \#9, \#10, \#11, \#12 \\
c\#13bridge            & \#13  \\
c\#18internal             & \#14, \#15, \#16, \#17 \\
c\#18bridge            & \#18  \\
etc             & \#5, \#6, \#7, \#8 \\
\hline
\end{tabular}
\end{center}
\end{table}


For walks without ARGEW, we set: number of walks per node = $20$, walk length = $10$, and context size = $3$. For walks with ARGEW, we set walks per node to $5$, and $low$ = $1$, $high$ = $9$ for rescaling.

\paragraph{Small $p$, large $q$ (inward walks)}
In this setting, walks tend to stay locally and avoid going outwards. It is known that such node2vec embeddings capture structural equivalence instead of homophily. Table~\ref{walkexperim_p1_q4} shows the proportion of coappearing nodes for each node type for the community of node \#13, when $p=1, q=4$. In particular, we examine the proportions of coappearing with the same community's internal nodes with the ``etc'' nodes.

In vanilla node2vec, the set of coappearing nodes for ``internal'' nodes is very different from that of the ``bridge'' node. Since the walks tend to stay inwards, ``internal'' nodes coappear more with the other ``internal'' nodes in the same community (around $70\%$), while the proportion of meeting ``etc'' nodes is below $1\%$. But the ``bridge'' nodes only meet their ``internal'' nodes around half of the time, and the proportion of coappearing with ``etc'' nodes is quite high: $35\%$. ``Bridge'' nodes are connected with the ``etc'' nodes nearly as much as with its ``internal'' nodes, and hence locally-moving walks tend to cover both types rather than being skewed to the ``internal'' nodes. Such a difference in the coappearance proportion will naturally result in ``internal'' nodes embedded far from the corresponding ``bridge'' node due to the principle that nodes with common co-occuring nodes in walks obtain similar node2vec embeddings.

However, with ARGEW, ``internal'' nodes and the ``bridge'' node have similar coappearance proportion distributions. In particular, for the ``bridge'' node, the proportion of meeting its``internal'' nodes increases considerably. This is because during ARGEW, ``internal'' nodes are replaced with their ``bridge'' node many times since each (``internal'', ``bridge'') node pair has a big edge weight. Consequently, for both ``internal'' and ``bridge'' types, the proportion of meeting ``internal'' nodes is near $80\%$. Such influence of ARGEW is directly shown by the decrease of the difference values in Table~\ref{walkexperim_p1_q4} compared to vanilla node2vec.  The similar coappearance proportions let the embeddings of ``internal'' nodes be close to their ``bridge'' node: that is, embeddings capture communities, or homophily, as desired.

The other two communities have the same resulting pattern, and moreover, the results are similar when $p=0.25, q=1$ since small $p$ and large $q$ both commonly affect the walks in such a way that they tend to stay locally.

\begin{table}[h]
\caption{Proportion of coappearing node types ($p=1, q=4$)} \label{walkexperim_p1_q4}
\begin{center}
\begin{tabular}{cccc}
\hline
Model         
& \begin{tabular}[c]{@{}c@{}} Node type \end{tabular} 
& \begin{tabular}[c]{@{}c@{}} with \\``c\#13internal'' \end{tabular}
& \begin{tabular}[c]{@{}c@{}} with \\ ``etc'' \end{tabular} \\
\hline \hline 
             & ``c\#13internal'' & $0.698$ & $0.083$ \\
node2vec             & ``c\#13bridge'' & $0.523$ & $0.345$ \\
\cline{2-4}
             & \textbf{difference} & $\mathbf{0.175}$ & $\mathbf{0.262}$ \\
\hline
             & ``c\#13internal'' & $0.790$ & $0.052$ \\
node2vec, ARGEW            & ``c\#13bridge'' & $0.802$ & $0.110$ \\
\cline{2-4}
            & \textbf{difference} & $\mathbf{0.012}$ & $\mathbf{0.058}$ \\
\hline
\end{tabular}
\end{center}
\end{table}


\begin{table}[h]
\caption{Proportion of coappearing node types ($p=1, q=0.25$)} \label{walkexperim_p1_q025}
\begin{center}
\begin{tabular}{cccc}
\hline
Model         
& \begin{tabular}[c]{@{}c@{}} Node type \end{tabular} 
& \begin{tabular}[c]{@{}c@{}} with \\``c\#13internal'' \end{tabular}
& \begin{tabular}[c]{@{}c@{}} with \\ ``etc'' \end{tabular} \\
\hline \hline 
             & ``c\#13internal'' & $0.528$ & $0.260$ \\
node2vec             & ``c\#13bridge'' & $0.454$ & $0.300$ \\
\cline{2-4}
             & \textbf{difference} & $\mathbf{0.074}$ & $\mathbf{0.040}$ \\
\hline
             & ``c\#13internal'' & $0.596$ & $0.199$ \\
node2vec, ARGEW            & ``c\#13bridge'' & $0.519$ & $0.285$ \\
\cline{2-4}
            & \textbf{difference} & $\mathbf{0.077}$ & $\mathbf{0.086}$ \\
\hline
\end{tabular}
\end{center}
\end{table}


\paragraph{Large $p$, small $q$ (outward walks)}
In contrast, this setting makes walks move further outwards, and embeddings become more like community detection. This is because both the ``bridge'' and its ``internal'' nodes will easily meet nodes outside the community, hence the coappearance distribution will not differ much. This is an easier setup for ARGEW since the original walks already have a favorable property for capturing homophily. Table~\ref{walkexperim_p1_q025} shows the coappearance proportions when $p=1, q=0.25$, again for the community with node \#13. As expected, node \#13 and its ``internal'' nodes have similar coappearance proportion distributions, both with and without ARGEW. This is shown by the fact that the difference values in Table~\ref{walkexperim_p1_q025} are very small for both models.

In summary, with a small $p$, large $q$ setting, ARGEW augmentation ``overwhelms'' the undesired property of the original walks, while in a large $p$, small $q$ setting, ARGEW ``reinforces'' the desired property of the walks, thereby ending up in similarly good embeddings regardless of the hyperparameters.

\section{Discussion and Conclusion}

In this paper, we proposed ARGEW, a  random walk augmentation technique that supports random walk based node embedding methods for homophilous graphs. 
ARGEW augments random walks in such a way that in the corpus, for nodes with strong edges, there are many sequences where the two are surrounded by the same set of neighbor nodes. It does so by replacing nodes in the sequences with the maximum-weight neighbors. To ensure that the new derived walks are perfectly realistic, only neighbors that also have edges with both the previous and the next nodes are considered.

Using two benchmark networks and one network based on user log data of an MMORPG, we demonstrated ARGEW achieves the desired pattern that node pairs with stronger edges tend to have closer embeddings. We also showed ARGEW brings performance gain in node classification with similarly good results as supervised GCN, even without any node attribute or label information during training. ARGEW is not just better overall but also not sensitive to parameter setting (i.e. consistently good), and this was explained by comparing the coappearance distribution with and without ARGEW. To the best of our knowledge, ARGEW is the first method that augments the random walks themselves in order to reflect edge weights to the node embeddings. Since ARGEW is independent of the random walk strategy, it can work with any random walk based node embedding method, and it does not have any constraint as long as the network is weighted and the walks are given, which are essentially the only assumptions. 

There are also some limitations and potential future work regarding our paper. First, since ARGEW derives additional walk subsequences, we need to reduce settings that are related to data size: number of walks per node and batch size. In our experiments, we chose the reduced values somewhat heuristically by simply considering our GPU memory limit. It would be interesting to thoroughly examine the optimal amount of reduction, especially in terms of the walk length. 
Another important issue is the focus on homophily vs structural equivalence. ARGEW's goal of making nodes with strong edges obtain similar embeddings is particularly useful for homophilous networks. However, one might instead want nodes with common structural roles to have closer embeddings. For example, in the graph in Figure~\ref{walksamplenetwork} with clear structural roles, one might want to embed ``bridge'' nodes closely to one another and embed ``internal'' nodes closely to one another. Hence it would be interesting to study how to augment random walks when the focus is structural equivalence instead of homophily.



\clearpage 
\bibliographystyle{ACM-Reference-Format}
\bibliography{sample-base}


\end{document}